\documentclass[aps,prb,floatfix,twocolumn,a4paper,referees]{revtex4}
\usepackage{amsfonts}
\usepackage{amssymb}
\usepackage{graphicx}

\begin{document}

\title{Delayed creation of entanglement in superconducting qubits interacting with a microwave field}

\author{M. Abdel-Aty}\email{abdelaty@hotmail.com}
\affiliation{Mathematics Department, College of Science, Bahrain
University, Bahrain \\ Laboratoire Collisions Agrgats Ractivit -
UMR5589, Universit Paul Sabatier, 31062 Toulouse Cedex 09, France }

\author{Mark Everitt}\email{m.j.everitt@physics.org}
\affiliation{Centre for Theoretical Physics, The British University,
El Sherouk City, Postal No. 11837, P.O. Box 43, Egypt\\ Department
of Physics, Loughborough University, Leicestershire, LE11 3TU, UK.
 }

\date{\today}

\begin{abstract}
We explore the role played by the intrinsic decoherence in superconducting
charge qubits in the presence of a microwave field applied as a magnetic flux.
We study how the delayed creation of entanglement, which is opposite to the
sudden death of entanglement, can be induced. We compute the time evolution of the population inversion, total
correlation and entanglement, taking into account the junction mixed state
and dissipation of the cavity field. We show that although decoherence
destroys the correlation of the junction and field, information of the
initial state may be obtained via quasi-probability distribution functions.
\end{abstract}



\maketitle

\section{Introduction}

Over the last decade, superconducting qubits have gained substantial
interest as devices for application in quantum information processing
\cite{you05,mak01}. Here, Josephson qubits are recognized as being
among the most promising devices to implement solid state quantum
computation \cite{ave00}. The manipulation of quantum states in
individual and coupled qubits (Cooper-pair box) has been
demonstrated experimentally in~[\onlinecite{mar02}] and the behavior of charge
oscillations in superconducting Cooper pair boxes weakly interacting
with an environment has been discussed in~[\onlinecite{ben08}]. Superconducting
circuits can behave like atoms and test quantum mechanics at
macroscopic scales and be used to conduct atomic-physics experiments
on a silicon chip \cite{you05}. Furthermore, the quantum dynamics of
a Cooper-pair box with a superconducting loop in the presence of a
non-classical microwave field have been investigated
in~[\onlinecite{you03}].

In order to utilize superconducting circuits as a resource for
quantum information processing the problem of engineering entangled
states in coupled systems must be addressed. One of the most
important problems under consideration is how to make a long-lived
and easily monitored entangled states within existing experimental
set-ups\cite{vid00}. At present, one of the main obstacles
in the development of a larger-scale solid state quantum logic
circuit is decoherence. It is therefore important to develop
strategies to minimize the effects of decoherence on the dynamics of
the qubit systems \cite{qas08,bel07}. Much of the work that has been
done in this field focuses on decoherence due to coupling to
environmental degrees of freedom. However, there exists another
intriguing possible source of fundamental decoherence that was
proposed by Milburn in~[\onlinecite{mil91a}]. This arises if,
instead of being a continuum, time is considered to progress with
some minimum, but finite, increment. Hence, the dynamics is not
governed by a single unitary evolution operator but rather some
stochastic sequence of identical incremental unitary
transformations. Taking this into account results in an equation of
motion resembling that of master equations for Markovian open
quantum systems. In other words, introducing a descritization of
time results in a fundamental decoherence that affects any quantum
system. It is therefore natural to ask what limits such intrinsic
decoherence would place on applications in quantum information
processing. Here we build on other work in this field\cite{intDec}
and study the delayed creation of entanglement in superconducting
qubits interacting with a microwave field.

The system that we study in this work closely resembles that of a qubit coupled to
 a quantum field mode which can be effectively represented using the Jaynes-Cummings
 model~\cite{jaycumm63,gerkni}. A recent study has indicated that this system can be
 used to understand the quantum to classical transition of a field mode when a suitable model
 of environmental decoherence is introduced~\cite{everitt09}. Hence, by observing the behaviour
 of the dynamics of the atomic inversion, the qubit can be used as a probe to estimate the amount
 of environmental dechorence to which the field is subjected. Milburn's model is slightly different
 in that decoherence arises because time is taken to increment, discontinuously, over short intervals.
 As the system does not evolve continuously, but under a stochastic sequence of
 identical unitary transformations, the associated decoherence effects the qubit and field equally.
 Nevertheless, it is our suggestion that systems that exhibit collapse and revival phenomena will
 provide a possible platform from which to test Milburn's proposal. The discontinuous evolution that
 is central to this model is independent of the system under consideration. Any test, or estimate of
 the time increment, should be obtained from a number of physically different sample systems which exhibit
 similar dynamical properties. Hence, collapse and revival forms good basis for testing intrinsic decoherence
 as this behaviour is seen in a number of different physical systems and is well
 understood.
In this paper our objective is to demonstrate proof of principal and show the effect that
intrinsic decoherence might have on a realistic system, a cooper pair box coupled to
a non-classical field (we have used experimental parameters from~[\onlinecite{leh03}]).
We note that due to the existence of other decohering effects it may not be practically possible to
 establish the validity (or not) of intrinsic decoherence if the time increments are too small.
 Even so, we believe that studying collapse and revival phenomena with carefully controlled environments
 will provide an upper bound to the size of the interval.


Specifically, we study the junction-field dynamics and their associated
entanglement properties. We demonstrate a protocol for entanglement engineering and characterization by
studying a special type of superconducting charge qubit, namely a
single Cooper-pair box coupled to a microwave field applied as a
magnetic flux. We pay careful attention to the crucial difference
between the strong or weak field regimes effect on the total
correlation of the junction-field system. In order to complete our
study from the perspective of a phase space approach we
discussed the Wigner quasi-probability distribution.
The present work is motivated by experimental results on Josephson
junction and normal metal flux qubits coupled to the environment
\cite{sch01}. We note that some theoretical discussions and analysis
of special cases of the problem at hand were given in Refs.
\onlinecite{you03,liu05,ste06} and experimental results were
predicted in Ref. \onlinecite{sch01}.

The organization of this paper is as follows: in section 2 we introduce the
model and formulate the master equation and present its time-dependent
analytical solution. In section 3, as an application, we employ the
analytical results obtained in section 2 to discuss the AC Josephson effect
and entanglement for different values of the intrinsic decoherence. In
section 4, we focus on the Wigner function which corresponds to the final
state of the charge-qubit system. Finally, we summarize the results in
section 5.

\begin{figure}[!tb]
\begin{center}
\resizebox*{0.32\textwidth}{!}{\includegraphics{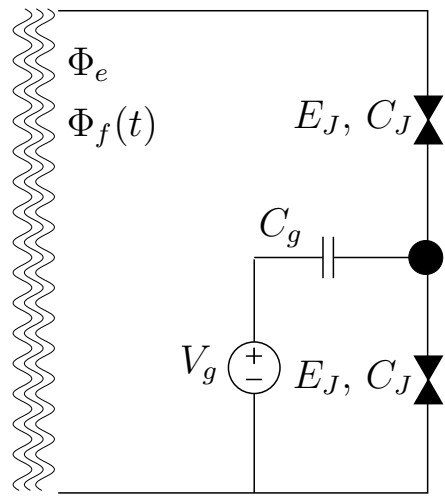}}
\end{center}
\caption{Schematic picture of the Cooper-pair box. The filled circle denotes
a superconducting island, the Cooper-pair box, which is biased by a voltage $%
V_{g}$ through the gate capacitance $C_{g}$ and coupled to the bulk
superconductors by two identical small Josephson junctions. The two
Josephson junctions have capacitance $C_{J}$ and Josephson energy $E_{J}$.
The total flux is the summation of the static magnetic flux $\Phi _{e}$, and
microwave-field-induced flux $\Phi _{f}(t)$ that applied via the
superconducting-quantum-interference-device loop.}
\end{figure}

\section{The model}

We consider a
superconducting box with a low-capacitance Josephson junction with the
capacitance $C_{J}$ and Josephson energy $E_{J}$, biased by a voltage source
$V_{g}$ through a gate capacitance $C_{g}$ which is externally controlled
and used to induce offset charges on the island. The schematic picture of
this single-qubit structure is shown in figure~1. The total
Hamiltonian of the system can then be written as \cite{mig01}
\begin{equation}
\hat{H}=\frac{(Q-C_{g}V_{g})^{2}e}{C_{g}+2C_{J}}-2E_{J0}\cos \phi
\cos \left( \frac{\pi \Phi}{\Phi _{0}}\right) +\hbar \omega (a
^{\dagger }a +\frac{1}{2}),  \nonumber  \label{5}
\end{equation}
where ${a }^{\dagger }$ and $a$ are, respectively, the creation and
annihilation operator of the cavity mode. In this structure, the
superconducting island with Cooper-pair charge $Q=2Ne$ is coupled to
a segment of a superconducting ring via two Josephson junctions,
where $e$ is the electron charge and $N$ is the number of
Cooper-pairs. We denote by $\phi =0.5(\phi _{1}+\phi _{2})$ the
phase difference across the junction. The gauge-invariant phase
drops $\phi _{1}$ and $\phi _{2}$ across the junctions are related
to the total flux $\Phi $ through the
superconducting-quantum-interference-device (SQUID) loop by the
constraint $\phi _{2}-\phi _{1}=2\pi \Phi /\Phi _{0},$ where $\Phi
_{0}=h/2e$ is the flux quantum. Here, $E_J=2E_{J0}\cos\phi\cos(\pi
\Phi/\Phi _{0})$ and self-inductance effects on the single-qubit
structure is ignored.

When a nonclassical microwave field is applied, the total flux is a
quantum variable $\Phi =\Phi _{e}+\Phi _{f}(t),$ where $\Phi _{f}$
is the microwave-field-induced flux. If we consider a planar cavity
and the SQUID loop of
the charge qubit is perpendicular to the cavity mirrors, the vector
potential of the nonclassical microwave field can be written as
$A(r)=|u_{\lambda }(r)|({a }^{\dagger }+{a }){A},$ where a
single-qubit structure is embedded in the microwave cavity with only
a single photon mode $\lambda$. Thus, the flux $\Phi _{f}$ can be
written as $\Phi _{f}=|\Phi _{\lambda }|({a }^{\dagger }+{a }),$
where $\Phi _{\lambda }=\oint u_{\lambda }.dl,$ where the contour
integration is over the interior of the SQUID loop. We shift the gate voltage
$V_{g}$ (and/or vary $\Phi _{e}$) to bring the single-qubit system
into resonance with $k$ photons: $E\approx k\hbar \omega _{\lambda
}$, $k=1,2,3,....$ It is to be noted that the charge states are not
the eigenstates of the Hamiltonian~(\ref{ham}), so that the
Hamiltonian can be diagonalized yielding the following two charge
states $|e\rangle =\cos \xi \left\vert 1\right\rangle -\sin \xi
\left\vert 0\right\rangle $ and $|g\rangle =\sin \xi \left\vert
1\right\rangle +\cos \xi \left\vert 0\right\rangle $ with $\xi
=\frac{1}{2}\tan ^{-1}(E_{J}/2\varepsilon ),$ where $\varepsilon
=2E_{c}(C_{g}V_{g}e^{-1}-(2n+1)).$ Employing these eigenstates to
represent the qubit, expanding the functions $\cos (\pi \Phi
_{e}/\Phi _{0})$ and using the rotating wave approximation, one can
derive the total Hamiltonian of the system as \cite{kre00}.
\begin{eqnarray}
\hat{H} &=\hbar\omega\left({n}+\frac{1}{2}\right) +&\left\{
\frac{1}{2}E-E_{J0}\sin \left( 2\xi \right) \cos \left( \frac{\pi
\Phi _{e}}{\Phi _{0}}\right) f({n})\right\} \sigma
_{z}  \nonumber \\
&&+\cos(2\xi)E_{J_0} \left\{a^{k}g_{(k)}(n)\sigma_{+}+ a^{\dagger
k}g_{(k)}^*(n)\sigma_{-}\right\}.
 \label{ham}
\end{eqnarray}
We denote by $\sigma _{\pm }$ and $\sigma _{z}$ the Pauli matrices in the
pseudo-spin basis and $g_{k}({n})$ represents the $k-$%
photon-mediated coupling between the charge qubit and the microwave field,
and are given by $g_{(1)}({n})=\sin \left( \frac{\pi \Phi _{e}}{\Phi
_{0}}\right) \left( \phi -\frac{1}{2!}\phi ^{3}{n}+\frac{1}{4!}\phi
^{5}(2{n}^{2}+1)-.....\right) ,$ $g_{(2)}({n})=\cos \left(
\frac{\pi \Phi _{e}}{\Phi _{0}}\right) (\frac{1}{2!}\phi ^{2}-\frac{2}{4!}%
\phi ^{4}(2{n}-1)$ $+.....,$ $g_{(3)}({n}))=\sin \left(
\frac{\pi \Phi _{e}}{\Phi _{0}}\right) \left( -\frac{1}{3!}\phi ^{3}+\frac{5%
}{5!}\phi ^{5}({n}-1)-.....\right) ,$ $etc.,$ $f({n})=\frac{1%
}{2!}\phi ^{2}(2{n}+1)-\frac{3}{4!}\phi ^{4}(2{n}^{2}+2%
{n}+1)+.....,$ ${n}=a ^{\dagger }a $ and $\phi =\pi
\frac{|\Phi _{\lambda }|}{\Phi _{0}}$.

We note that this model is similar to that of the Jaynes-Cummings
Hamiltonian~\cite{jaycumm63,gerkni}, the difference arising in
the coupling and intensity dependent term $E_{J0}\sin \left( 2\xi
\right) \cos \left( \frac{\pi \Phi _{e}}{\Phi _{0}}\right) f({n})$.
It is well known that the Jaynes-Cummings model exhibits collapse
and revival phenomena. That is, although the qubit initially
exhibits Rabi like oscillations these apparently disappear and then
subsequently revive~\cite{ebe80,naro81,kni82a,gerkni}.  This
phenomena  is well understood both
theoretically~\cite{geaban90,geaban91,knibuz95} and
experimentally~\cite{auf03,meu05}. Hence, we might expect this
system to behave in a similar fashion. As we will see, this is
indeed the case.

\subsection{Solution}

The intrinsic decoherence approach \cite{mil91a,mil91b,bre02,lid03,gar00}, is based
on the assumption that on sufficiently short-time steps the system does not
evolve continuously under unitary evolution but rather in a stochastic
sequence of identical unitary transformations. Under the Markovian
approximation \cite{mil91a,mil91b}, the master equation governing the time evolution
for the system is given, to first order in $\gamma^{-1}$, by
\begin{equation}
\frac{d}{dt}\hat{\rho}(t)=-\frac{i}{\hbar }[\hat{H},\hat{\rho}]-\frac{\gamma
}{2\hbar ^{2}}[\hat{H},[\hat{H},\hat{\rho}]],  \label{mas}
\end{equation}
where $\gamma$ is the intrinsic decoherence parameter. The first term on
the right-hand side of equation (\ref{mas}) generates a coherent unitary
evolution of the density matrix, while the second term represents the
decoherence effect on the system and generates an incoherent dynamics of the
qubits system. In order to obtain an exact solution for the density operator
$\hat{\rho}(t)$ of the master equation (\ref{mas}), three auxiliary
superoperators ${J},\ {S}$ and ${L}$ can be introduced
\cite{mil91a,mil91b} as ${J}\hat{\rho}(t)=\gamma \hat{H}\hat{\rho}(t)\hat{H}$,
${S}\hat{\rho}(t)=-i[\hat{H},\hat{\rho}(t)]$ and
${L}\hat{\rho}(t)=-(\gamma /2)(\hat{H}^{2}\hat{\rho}(t)+\hat{\rho}(t)\hat{H}^{2})$.

Unlike state vectors, the density matrix can represent statistical
mixtures of states. Hence, intrinsic decoherence implies that pure
states may evolve into mixed states. It is intriguing to ask a
related question; if some component of a system is initially in a
mixed state is it possible, in the presence of intrinsic
decoherence, to use this state to develop resources for quantum
information processing. In order to prepare such an initial state we
might consider the deliberate, and temporary, application of an
appropriate decohering environment. Alternatively, it may be
possible to apply some form of guidance law~\cite{Ralph04} to attain
the desired state within the Bloch sphere. Here, we suppose that the
initial state of the Cooper pair box is given by $\rho
_{A}(0)=\varsigma _{1}|e\rangle \langle e|+\varsigma _{2}|g\rangle
\langle g|$, where $\varsigma _{i}\geq 0,$ and $\varsigma
_{1}+\varsigma _{2}=1$. In addition, we suppose that the initial
state of the field is given by $\rho _{F}(0)=|\varpi \rangle \langle
\varpi |,$ where $|\varpi \rangle =\sum\limits_{n=0}^{\infty
}b_{n}|n\rangle ,$ and $b_{n}^{2}=|\langle \varpi |n\rangle |^{2}$
being the probability distribution of photon number for the initial
state. It is then straightforward to write down the formal solution
of the master equation (\ref{mas}) as follows
\begin{widetext}
\begin{eqnarray}
\hat{\rho}(t) &=&\exp ({J}t)\exp ({S}t)\exp ({L}t)%
\hat{\rho}(0)
\nonumber
\\
&=&
\sum_{k=0}^{\infty }\left\{ \varsigma _{1}\hat{M}_{k}(t)|e,\varpi
\rangle
\langle \varpi ,e|\hat{M}_{k}^{\dagger }(t)+\varsigma _{2}\hat{M}%
_{k}(t)|g,\varpi \rangle \langle \varpi ,g|\hat{M}_{k}^{\dagger
}(t)\right\} ,  \label{dens}
\end{eqnarray}
\end{widetext}
where $\hat{\rho}(0)$ is the density operator of the initial state of the
system and
\begin{equation}
\hat{M}_{k}=\frac{\left( \gamma t\right) ^{k/2}}{\sqrt{k!}}\hat{H}^{k}\exp
\left( -i\hat{H}t\right) \exp \left( -\frac{\gamma t}{2}\hat{H}^{2}\right) .
\end{equation}%
The so-called Kraus operators $\hat{M}_{k}$ satisfy $\sum\limits_{k=0}^{%
\infty }\hat{M}_{k}(t)\hat{M}_{k}^{\dagger }(t)=\hat{I}$ for all $t$.

\section{Dynamics}

\subsection{AC Josephson effect}
The AC Josephson effect can be used to observe collapse and revival
in a condensed matter system. AC Josephson effect involve
interaction of the photons with a junction which behaves like an
atom undergoing transition between the quantum states of each side
of the junction as it adsorbs and emits radiation. To see this
phenomenon, we calculate the the time dependence of the populations
inversion $\langle \hat{\sigma}_{z}(t)\rangle$ for different values
of the decoherence parameter $\gamma $.

\begin{figure}[!t]
\begin{center}
\resizebox*{0.48\textwidth}{!}{\includegraphics{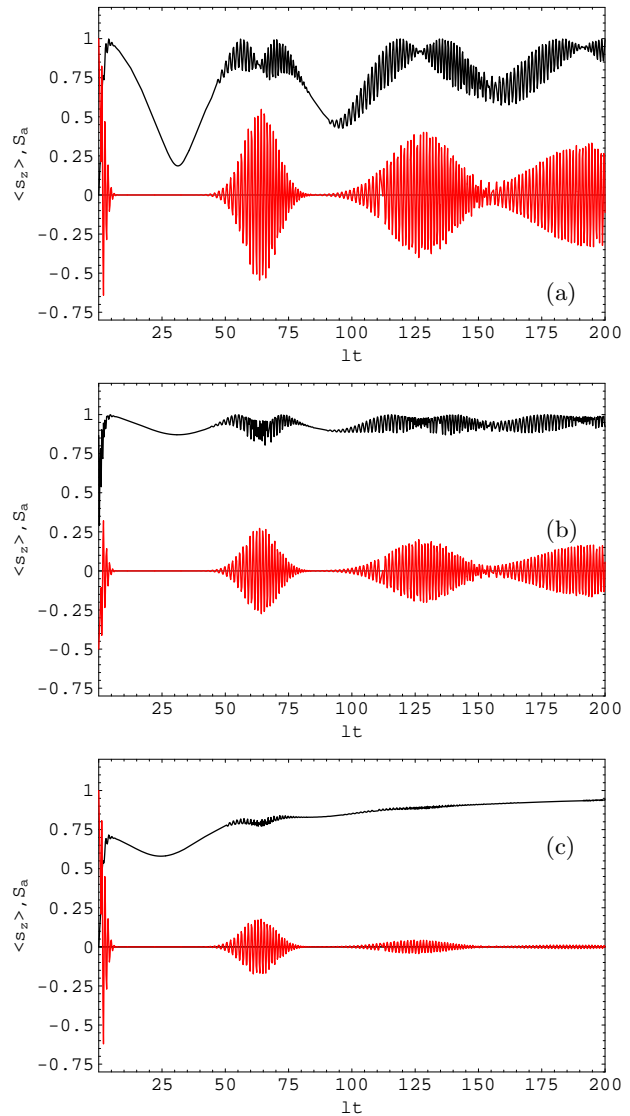}}
\put(-35,115){(c)}\put(-35,235){(b)}\put(-35,325){(a)}
\end{center}
\caption{(color online) Dynamics of the atomic inversion and tangle
as functions of the scaled time $\lambda t$. The parameters are
$\varsigma_1 = cos^2\theta $ and $\bar{n} = 25$. The other
parameters are $\phi = 0.1$; $\xi = \pi/2$; $E_{J 0} = 1$ and $k =
1$. The three figures from above have (a) $\theta=0$, $\gamma=0$,
(b) $\theta=\pi/3$, $\gamma=0$, and (c) $\theta=0$, $\gamma=0.001$.}
\end{figure}

In Fig. 2, we plot the populations inversion and tangle as a
function of the scaled time $\lambda t$ where
$\lambda=\sqrt{e^2\omega/\hbar C_F}$. In this figure, we use the
generalized mixed state form of the tangle~\cite{run01}
$S_a=2\min_{S_i}\sum_{i}(1-Tr[(\rho_a^{(i)})^2])$, where
$\rho_a^{(i)}$ is the marginal state for the $i^\mathrm{th}$ term in the
ensemble decomposition and $S_i$ is a convex combination of the pure
states. In the pure state case this definition will be reduced to
the usual from of the tangle $S_a=2(1-Tr\rho_a^2)$. Here we consider
the single photon process and $\Phi _{e}=\Phi_0/2$ for all Josephson
charge qubits, and the gate voltage is adjusted to have the qubit on
resonance with the cavity mode. It is interesting to show collapse
and revival  of the populations inversion as the system evolves,
from an initial coherent state, for the microwave field.
The interesting feature of the present case is that the
general state without decoherence effect is host to entanglement
between the junction and the field and this provides an opportunity to study
the dynamics of such entanglement.
To throw further light on the matter we also
show, in Fig. 2, the time evolution of the mixed state tangle calculated
from $\hat{\rho}(t)$ for the mixed initial state.
Evidently, just as the occupation of the initial qubit states
collapses and revives, so does the tangle. In close analogy with the
Jaynes-Cummings model, the tangle remains near zero for long periods
between revivals. The deterioration of revivals may be seen for
$\gamma /\lambda =0.01$ where the amplitude of oscillations is
smaller than in the case for larger $\gamma /\lambda $ where the
revivales do not occur. Once the initial state of the field is
considered to be a thermal state, the collapse-revival structure
disappears. This initial atomic inversion collapses to the mid-level
and does not oscillate for some period of time, which is followed by
a quasichaotic behavior (in agreement with the standard two-level
system \cite{kni82}).

It is evident that the effect of the mixed state parameter
$\theta=\pi/3$ leads to decreasing of
$\langle\hat\sigma_z(t)\rangle$ amplitude and increasing of the
local maximum of the tangle with small oscillations (see Fig. 2b).
These small oscillations will be observed even for a large
interaction time. If we consider $\theta=\pi/4$, i.e. the initial
state of the Cooper pair box is given by  $\rho
_{A}(0)=0.5(|e\rangle \langle e|+|g\rangle \langle g|)$, we see that
$\langle\hat\sigma_z(t)\rangle=0$ and the tangle tends to its
maximum value. In this sense, the system has relatively high
entanglement. The smallest entanglement which is almost zero is
obtained only for initial pure state case. Similarly to what we have
seen in Fig. 2b, we observe how the amplitude of
$\langle\hat\sigma_z(t)\rangle$ decreases as the time goes on due to
the inclusion of intrinsic decoherence (see Fig. 2c). Recall that,
as time increases further, the intrinsic decoherence effect leads to
zero inversion and maximum entanglement.

It is interesting to mention here the fact that, in the absence of the
decoherence, the final density matrix of the system Eq. (\ref{dens}) can be
rewritten as $\hat{\rho}(t)=$ $|\psi (t)\rangle \langle \psi (t)|$ where $%
|\psi (t)\rangle =|e\rangle \otimes |\Phi _{e}\rangle +|g\rangle \otimes
|\Phi _{g}\rangle $ and $|\Phi _{j}\rangle $ are the field states \cite%
{ger05}. From our further calculations, we show that, at a particular
effective time the populations inversion and the overlap of field states
vanish simultaneously. That is, the populations become equal and both the
Cooper pair box and field states are mutually orthogonal. Thereby, the state
equation $|\psi (t)\rangle $ becomes maximally entangled at this particular
effective time.  We would like to remark that the interest in studying the
generation of maximally entangled states in the short time region arises due
to the unavoidable presence of decoherence effects in experiments performed
at the high $Q$ region due to cavity dissipation. The experimental generation
of entangled states in the mesoscopic field regime has been reported for
fields with an average photon number of a few tens \cite{auf03}. In this
regime, after the first collapse, the field splits into two orthogonal humps
in phase space. Thus, the field behaves as an effective two-level system and
the entanglement of the composite system is easily described in terms of
factorized states \cite{ban05,ret06}.

\subsection{Relative entropy}

For the entangled states $\hat{\rho}(t)$ the quantum relative entropy is
defined as the distance between the entangled state $\hat{\rho}(t)$ and
disentangled state $tr_{\mathcal{A}}\hat{\rho}(t)\otimes tr_{\mathcal{B}}%
\hat{\rho}(t)\in \mathfrak{S}(\mathcal{H}_{1}\otimes \mathcal{H}_{2})$ \cite%
{aty01}
\begin{equation}
I_{\rho }\left( \rho _{t}^{\mathcal{A}},\rho _{t}^{\mathcal{B}}\right) =tr%
\hat{\rho}(t)(\log \hat{\rho}(t)-\log (tr_{\mathcal{A}}\hat{\rho}(t)\otimes
tr_{_{\mathcal{B}}}\hat{\rho}(t))),  \label{me}
\end{equation}%
where $\rho _{t}^{\mathcal{A}}=tr_{\mathcal{B}}\left( \hat{\rho}(t)\right) $
and $\rho _{t}^{\mathcal{B}}=tr_{\mathcal{A}}\left( \hat{\rho}(t)\right) ,$ $%
\mathcal{A(B)}$ refers to the first (second) qubit$.$ Note that if the
entangled state $\hat{\rho}(t)$ is a pure state, $S(\hat{\rho}(t))=0$ and
then $S(tr_{\mathcal{A}}\hat{\rho}(t))=S(tr_{\mathcal{B}}\hat{\rho}(t)),$
which means that we have $I_{\rho }\left( \rho _{t}^{A},\rho _{t}^{B}\right)
=2S(tr_{\mathcal{B}}\hat{\rho}(t))$.

Taking the partial trace over the junction system, we obtain $\rho
_{t}^{F}=tr_{A}\rho (t).$ The von Neumann entropy for the reduced state $%
S(\rho _{t}^{F})$ is computed as
\begin{eqnarray}
S(\rho _{t}^{F})&=&-\lambda _{1}^{F}(t)\log \lambda _{1}^{F}(t)-\lambda
_{2}^{F}(t)\log \lambda _{2}^{F}(t)\nonumber\\&&-\lambda _{3}^{F}(t)\log \lambda
_{3}^{F}(t)-\lambda _{4}^{F}(t)\log \lambda _{4}^{F}(t),  \label{fe}
\end{eqnarray}%
where $\lambda _{i}^{F}(t)$ are the solutions of the following equation $%
\det [\hat{\rho (t)}-\lambda (t)\hat{N(t)}]=0,$ $\hat{\rho (t)}$ and $\hat{%
N(t)}$ are $4\times 4$ matrices having the following elements
\begin{eqnarray}
\left[ \hat{\rho (t)}\right] _{ij} &\equiv &\langle \psi _{i}(t)|\rho
_{t}^{F}|\psi _{j}(t)\rangle ,\quad (i,j=1,2,3,4),  \nonumber \\
\left[ \hat{N(t)}\right] _{ij} &\equiv &\langle \psi _{i}(t)|\psi
_{j}(t)\rangle ,\quad (i,j=1,2,3,4),
\end{eqnarray}%
and $|\psi _{j}(t)\rangle $ are the eigenfunctions of the following
eigenvalue problem $\rho _{t}^{F}|\psi _{i}(t)\rangle =\lambda
_{i}^{F}(t)|\psi _{i}(t)\rangle .$

On the other hand, the final state of the junction system is given by taking
the partial trace over the field system $\rho _{t}^{J}\equiv tr_{F}\rho (t).$
Then the von Neumann entropy for the reduced state $S(\rho _{t}^{A})$ is
computed by
\begin{equation}
S(\rho _{t}^{J})=-\lambda _{+}^{J}(t)\log \lambda _{+}^{J}(t)-\lambda
_{-}^{J}(t)\log \lambda _{-}^{J}(t),  \label{ae}
\end{equation}%
where $\lambda _{\pm }^{J}(t)$ are the eigenvalues of the reduced junction
state $\rho _{t}^{J}.$

Using equations (\ref{me}), (\ref{fe}) and (\ref{ae}), we
obtain the relative entropy, which can be used to measure the total
correlation in the system under consideration. From the above equations and
the matrix elements which represent the state of the field, we are able to
determine under which conditions we may attain reasonable correlations
between the junction and cavity field. Apart from the case represented
above, in general there is no way to relate the correlation dynamics
exclusively to the initial field state.

\begin{figure}[!t]
\begin{center}
\resizebox*{0.48\textwidth}{!}{\includegraphics{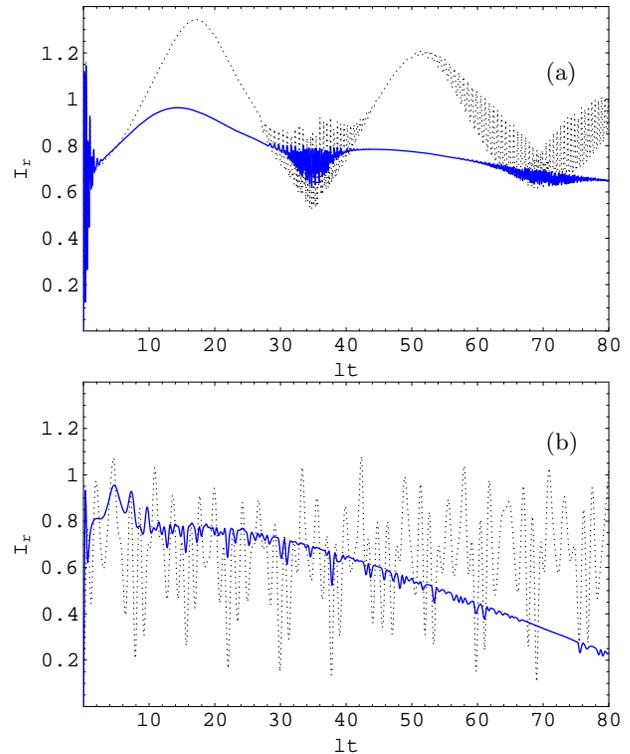}}
\put(-35,125){(b)}\put(-35,265){(a)}
\end{center}
\caption{(color online) Plot of $I_{\protect\rho }\left( \protect\rho _{t}^{\mathcal{A}},%
\protect\rho _{t}^{\mathcal{B}}\right) $ as a function of the scaled time $%
\protect\lambda t.$ The Cooper pair box starts from a mixed state with $%
\protect\varsigma _{1}=0.9$ and the initial state of the field is a coherent
state with different values of the mean photon number, where (a) $\overline{n%
}=30$ and (b) $\overline{n}=0.5$. The solid curve corresponds to presence of
the decoherence ($\protect\gamma =0.1\protect\lambda $) and dotted curve
corresponds to absence of the decoherence ($\protect\gamma =0).$}
\label{dem1}
\end{figure}

In what follows, we shall analyze numerically whether and how it would be
possible to control the correlations of the output state. In typical
experiments, the Cooper-pair box can be made from aluminum, with an energy
gap of $2.4K$ (about $50$ GHz) \cite{leh03}, the charge energy $149$ GHz and
the Josephson energy $13.0$ GHz. The frequency of the cavity field is taken
as $40$ GHz, corresponding to a wavelength $\sim 0.75$ cm. As a side remark,
we note that, the value $I_{\rho }\left( \rho _{t}^{\mathcal{A}},\rho _{t}^{%
\mathcal{B}}\right) =2\ln (2)$ appears in analytical criterion, based on an
estimation of the composite system entanglement. It will be worth analyzing
this curious coincidence in order to gain a deeper understanding of the
strengths and maximum points of the correlations. The points of maximum
correlation can be generated either using equation (\ref{me}) or, what seems
more feasible, directly from Fig. (\ref{dem1}). In this figure, we plot $%
I_{\rho }\left( \rho _{t}^{\mathcal{A}},\rho _{t}^{\mathcal{B}}\right) $ as
a function of the scaled time and for different values of the decoherence
parameter $\gamma $, which is in agreement (at $t=0$) with the exact
calculations for the $I_{\rho }\left( \rho _{t}^{\mathcal{A}},\rho _{t}^{%
\mathcal{B}}\right) $. This figure clearly shows that the correlations
reaches a local maximum before the first collapse time. Returning to Fig.~\ref{dem1},
after reaching its maximum the correlation decays exponentially
as a result of the dissipation and/or the microwave field, attaining local
minimum in the asymptotic limit. Furthermore, the addition of intrinsic
decoherence via the parameter $\gamma $ leads to a drastic change of the
correlation properties, lowering the maximum values of the total
correlations. It is interesting to note that in this regime, the maximum
correlations occur at a time in which the two orthogonal states are
completely overlapped in phase space.

Since maximum correlation takes a finite time, it makes sense to consider
whether the small values of the mean photon number can be used to change the
process of correlations. It is shown that, a similar effect of the
decoherence on the total correlation is seen in Fig. \ref{dem1}b for a small
value of the mean-photon number, but the number oscillations is increased
with decreasing the local minimum $I_{\rho }\left( \rho _{t}^{\mathcal{A}%
},\rho _{t}^{\mathcal{B}}\right) \simeq 0.2$. We attribute our finding to
the fact that in the presence of decoherence it takes some time for the
junction to become entangled with the microwave field. If during this
entanglement buildup the system is acted upon by an appropriately designed
control, it becomes possible to channel back quantum coherence from the
field to the Cooper pair box. Our results thus reinforce the superconducting
charge qubits as viable candidates for quantum-information processing
devices, and suggest that more refined control strategies might play an
important role in future of solid-state based quantum information devices.

From the above results, one might now raise the following possibility:
taking a strong decoherence regime where $\gamma \gg 0.01\lambda $, one can
estimate the exact time in which the total correlations will be vanished.
Therefore, if the intrinsic decoherence is substantially strong, no
correlations between the junction and the microwave field will occur,
suggesting that it might be possible to obtain maximum entangled states
using a weak decoherence. It is straightforward to verify, however, that in
order to do so one would require a value of $\gamma $ close to zero.

\subsection{Concurrence}

To measure the degree of entanglement for mixed states of bipartite systems
composed by two-level subsystems, one needs to consider a commonly used
measure such as the concurrence \cite{min05} which has been proven to be a
reasonable entanglement measure or negativity \cite{vid02}. Analysis of the
entanglement decay rates under decoherence for different models of the
interaction between systems of arbitrary dimensions with the environment has
been presented \cite{car07}. We write the standard basis of the product
space of the system as $|00\rangle =$ $|g,n\rangle ,$ $|01\rangle =$ $%
|e,n\rangle ,$ $|10\rangle =$ $|g,n+1\rangle $ and $|11\rangle =$ $%
|e,n+1\rangle .$ For the density matrix $\hat{\rho}(t),$ which represents
the state of a bipartite system, concurrence is defined as
\begin{equation}
C(\hat{\rho})=\max \{0,\lambda _{1}-\lambda _{2}-\lambda _{3}-\lambda _{4}\},
\label{coc}
\end{equation}%
where the $\lambda _{i}$ are the non-negative eigenvalues, in decreasing
order ($\lambda _{1}\geq \lambda _{2}\geq \lambda _{3}\geq \lambda _{4}$),
of the Hermitian matrix ${\Upsilon }\equiv \sqrt{\sqrt{\hat{\rho}}%
\widetilde{\rho }\sqrt{\hat{\rho}}}$ and $\widetilde{\rho }=\left( {%
\sigma }_{y}\otimes {\sigma }_{y}\right) \hat{\rho}^{\ast }\left(
{\sigma }_{y}\otimes {\sigma }_{y}\right) $. Here, $\hat{\rho%
}^{\ast }$ represents the complex conjugate of the density matrix $\hat{\rho}
$ when it is expressed in a fixed basis and ${\sigma }_{y}$
represents the Pauli matrix in the same basis. The function $C(\hat{\rho})$
ranges from $0$ for a separable state to $1$ for a maximum entanglement.

\begin{figure}[!t]
\begin{center}
\resizebox*{0.48\textwidth}{!}{\includegraphics{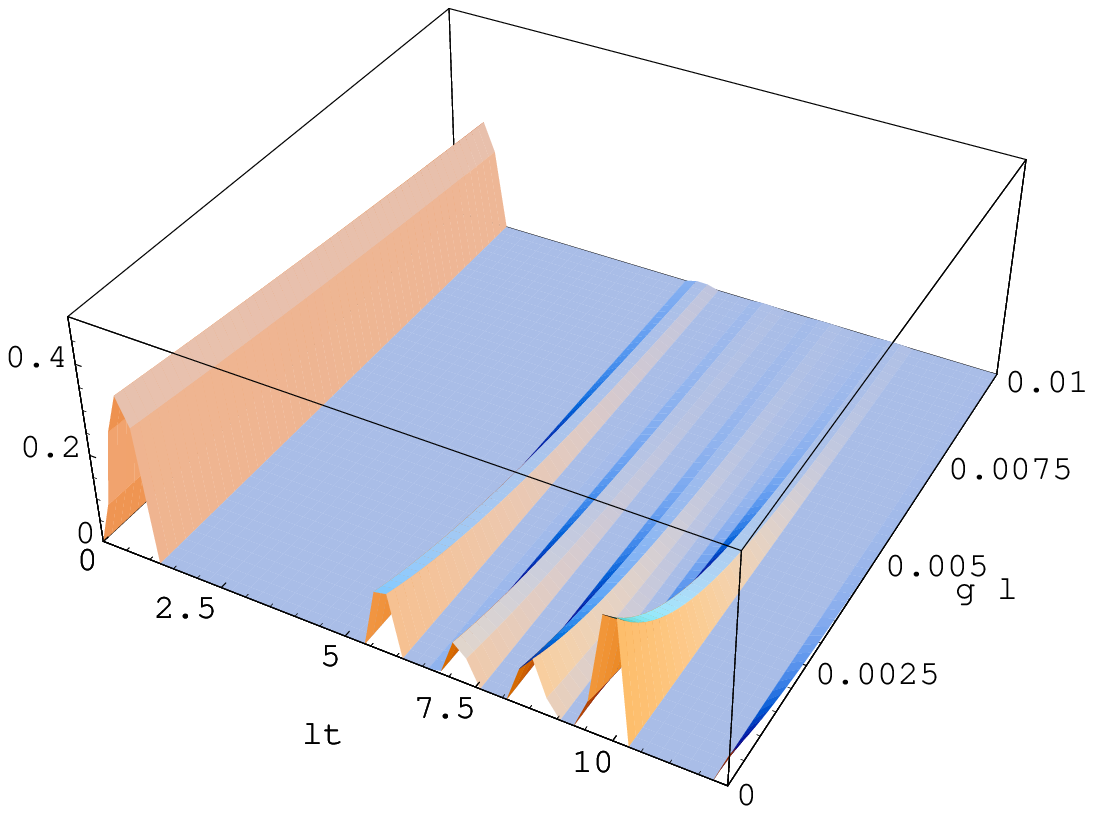}}
\put(-60,135){(a)} \\%
\resizebox*{0.48\textwidth}{!}{\includegraphics{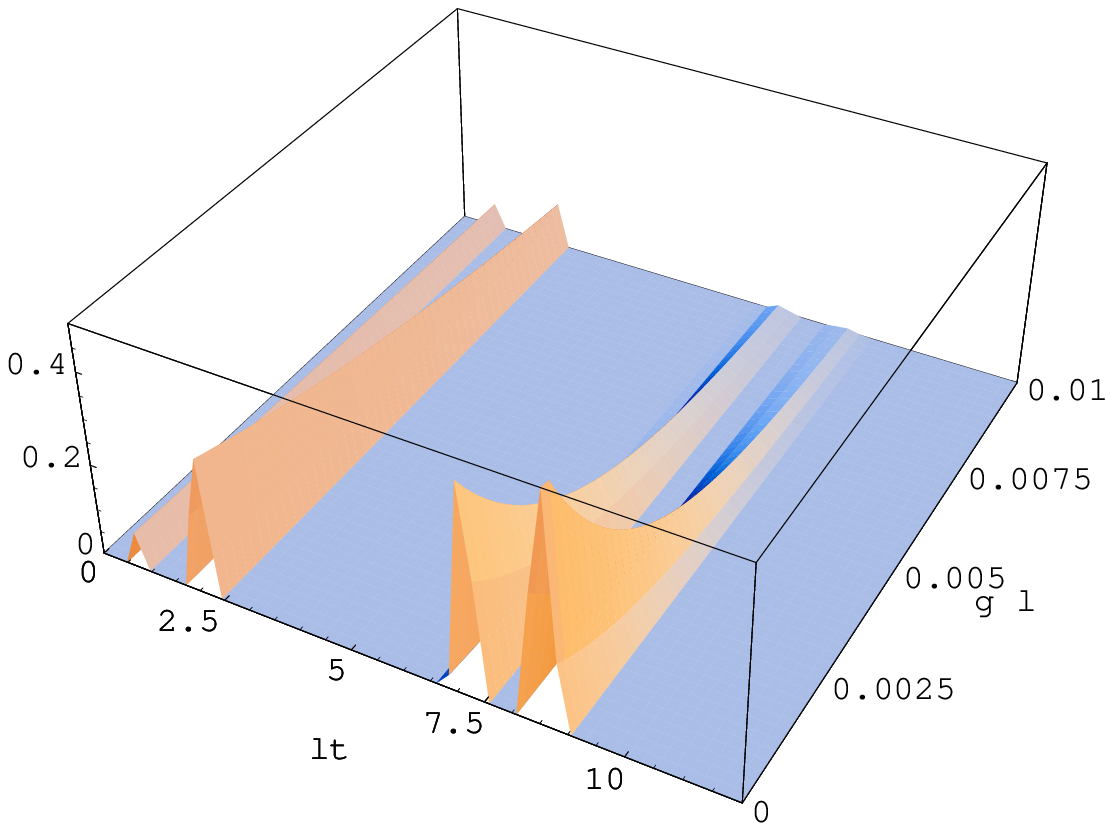}}
\put(-60,135){(b)}
\end{center}
\caption{(color online) Development of the concurrence $C({\protect\rho })$ as a
function of the scaled time $\protect\lambda t$ and the decoherence
parameter $\protect\gamma /\protect\lambda $. The other parameters are the
same as figure 3. }
\label{conc}
\end{figure}

In figure (\ref{conc}), we plot the numerically evaluated results
for the concurrence $C(\hat{\rho})$, as a function of the scaled
time $\lambda t$ and phase damping parameter $\gamma $ in units of
$\lambda $. The overall picture coming from Fig. (\ref{conc}) is
that the recurrence dynamics due to large mean photon number
($\overline{n}=30$) starts from zero, goes to a maximum and then
returns to zero. In an asymptotic limit the coupled system reaches a
pure state and stays in this pure state some times ($1\leq \lambda
t\leq 5.3$) then at some finite time weak entanglement is observed.
Meanwhile, as shown in Fig. (\ref{conc}b), due to small value of the
mean photon number, no entanglement at earlier times, and suddenly
at a certain time entanglement starts to build up. This delayed
creation of entanglement, that has been called sudden birth of
entanglement \cite{fic06}, is opposite to the currently extensively
discussed sudden death of entanglement \cite{yu07}. We see that
$C(\hat{\rho})$ falls off sharply as $\lambda t$ increases. The
non-classical character of the field for small values of the average
photon number $\overline{n}$, is reflected in larger entanglement
between the junction and the microwave field. As long as the time is
reasonably close to this threshold value and the mean photon number
takes small values, then the delayed creation of entanglement is
still obtained. However, if we start from a pure state, the
asymptotic value of the concurrence can be obtained when the phase
damping is increased. It is observed that, at later times
decoherence destroys such structures, in agreement with Fig.
\ref{dem1}. We have confirmed the predictions of this phenomenon
using a systematic numerical analysis where a number of relevant
parameters has been varied. However, once the initial state setting
of the Cooper pair box is considered as a mixed state, this feature
no longer exists and entanglement vanishes in an asymptotic limit
(see Fig. \ref{conc}).

\section{Phase space approach}

To give a more detailed discussion, we are now going to focus our attention on the
field dynamics. The representation of fields in phase space has been
providing new insights of the two-level system dynamics \cite{moy06}.
Perhaps the most convenient quasiprobability to be used in this kind of
problem is the $W-$function. In this connection, it seems interesting to
consider the Wigner function corresponding to the final state of the present
system. Experimentally these functions can be measured via homodyne
tomography~\cite{yur86}. Since the quasi-probability distribution function
is the Fourier transform of the characteristic function, therefore the
Wigner function can be obtained from the direct evaluation of the integral
\begin{equation}
W(x,p)=\frac{1}{\pi }\int\limits_{-\infty }^{\infty }\langle x-x^{\prime
}|\rho (t)|x+x^{\prime }\rangle \exp \left( 2ipx^{\prime }\right) dx^{\prime
},  \label{wf}
\end{equation}
\begin{figure}[t]
\begin{center}
\resizebox*{0.48\textwidth}{!}{\includegraphics{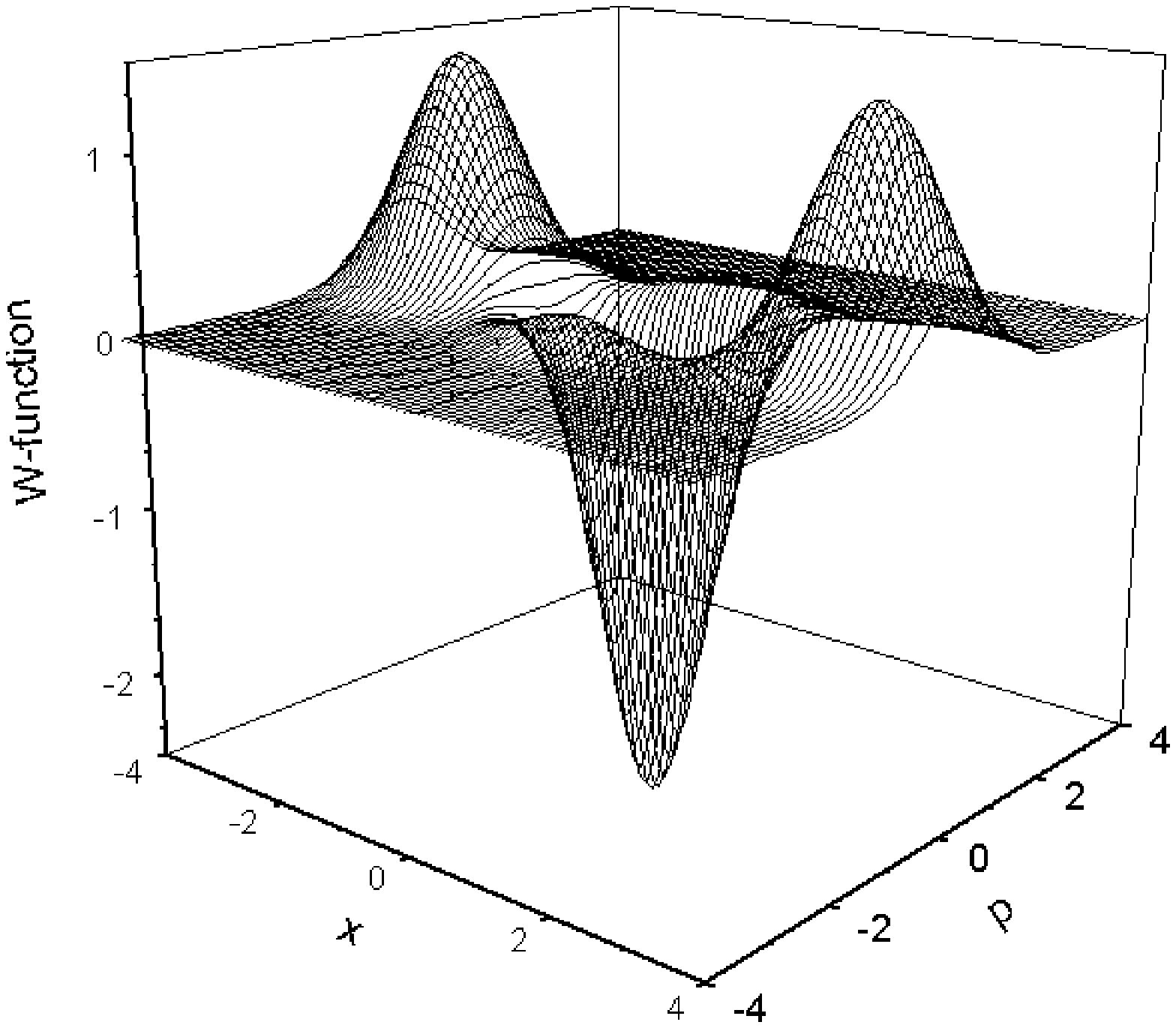}}
\put(-30,175){(a)} \
\resizebox*{0.48\textwidth}{!}{\includegraphics{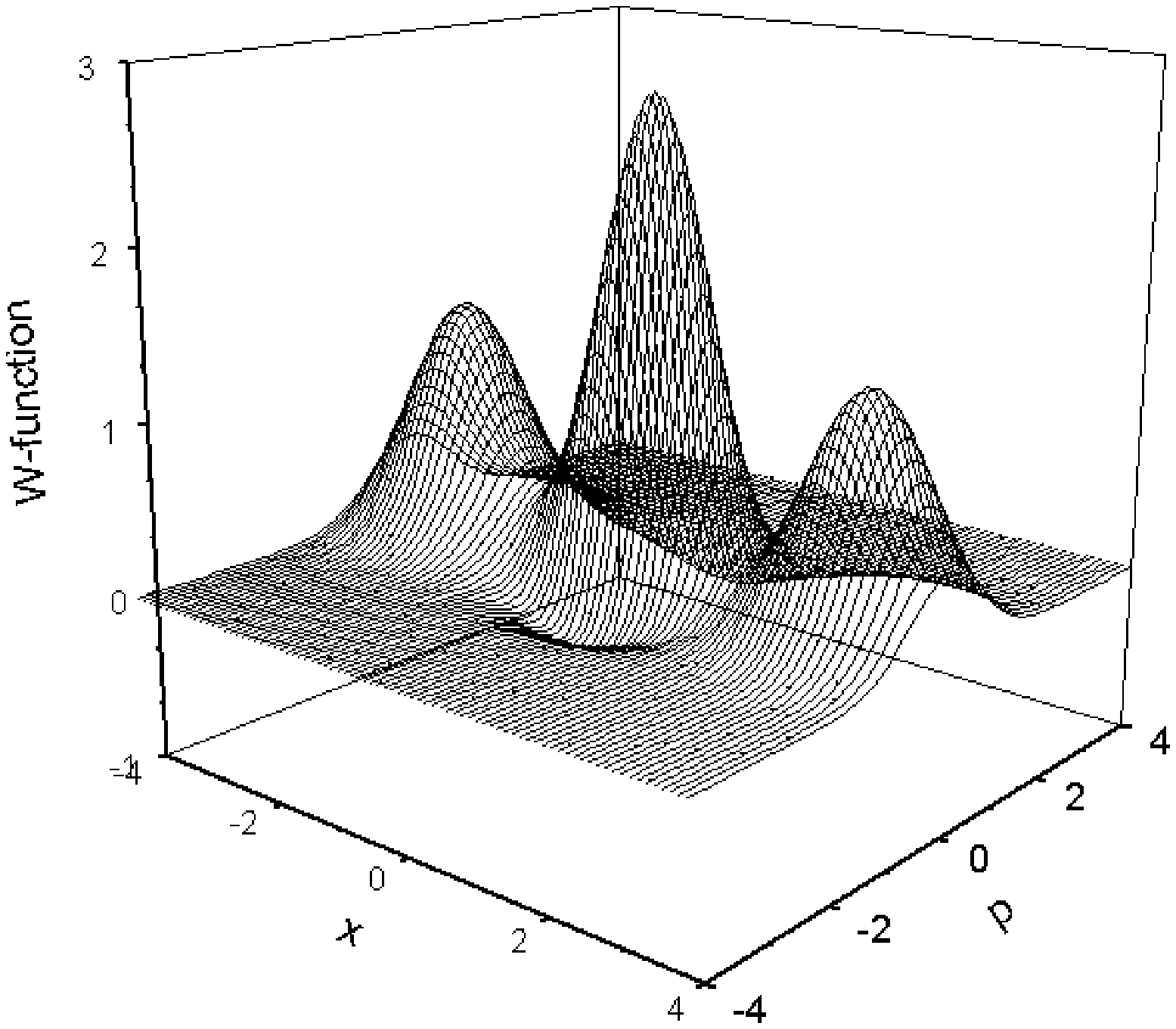}}
 \put(-30,175){(b)}
\end{center}
\caption{Wigner distribution of the cavity field as function of $x$
and $p$. The parameters are $\protect\alpha =2.1,t=0,\protect\theta
=\protect\pi /2$ , (b) $\protect\alpha =2.1,t=0,\protect\theta =0.$
In this case we have
assumed the initial state of the junction is given by $|\protect\psi (0)
\rangle _{J}=\cos (\protect\theta /2)|e\rangle +\sin (\protect\theta
/2)|g\rangle .$  } \label{w1}
\end{figure}
According to Eq. (\ref{wf}) we have plotted in Fig. \ref{w1}, the
Wigner function $W(x,p)$ against the parameters $x$ and $p$ $\ $for
the initial state of the filed as coherent state. If the Cooper pair
box starts from a superposition state ($\theta=\pi/2$), we see that
the Wigner function has negative values with a very little structure
around the base, see Fig. \ref{wf}a. This in fact is a signature of
the non-classical effect. These negative values will not be seen if
the Cooper pair box starts from a pure state Fig. \ref{wf}b. From
our further calculations, one may see that due to the decoherence
the quasiprobability distribution function is not as negative as in
absence of the decoherence. Of course, for greater values of the
decadence parameter the effect would be stronger. For $t>0$ the two
peaks split into two sets of counter-rotating peaks during the
collapse. At longer times the $W-$function is spread out over an
angular region in the $xy-$plane. If we combine this observation
with the fact that the entanglement degree at this moment is almost
maximum we can conclude that the cavity field is in a pure state. As
shown in \cite{moy06} although decoherence destroys the quantumness
of the field, information of the initial field may be obtained via
the reconstruction of quasiprobability distribution functions.

\section{Concluding remarks}

We have extended the exactly solvable model of a single-Cooper-pair box with
a nonclassical microwave field, taking into account the decoherence effect.
We considered the charge-qubit with a
superconducting-quantum-interference-device loop and used the microwave
field to change the flux through the loop. This treatment puts the so far
phenomenological description of charge-qubit systems on a firm footing and
paves the way for a variety of future applications in particular because the
extension to an arbitrary number of charge qubits is straightforward.
Collapse and revival phenomenon have been predicted via the Josephson effect
that involve interactions of the photons with the junction. The revivals
which may be seen in population inversion for small values of the
decoherence parameter will not be occurred for stronger decoherence. It is
shown that, for certain values of the initial average photon number,
maximally entangled state is generated.

We have presented a detailed analysis of the dynamics of the
process of the total correlation in a junction-field system. For the
weak-field region, the behavior is essentially oscillatory, while for the
strong-field region, the correlation tends to be randomized in the
evolution, although showing collapses and revivals. To measure the degree of
entanglement for mixed states of junction-field system, we have considered
the concurrence which is a commonly used measure. We have investigated numerically
results of the concurrence by considering the influences of the mean-photon
number and intrinsic decoherence in the junction-field system. We have shown
an interesting phenomenon of delayed birth of entanglement that initially
separable junction and field become entangled after a finite time. Similar
to the correlation case, the decoherence leads to lowering the maximum
entanglement and further increase of the decoherence leads to
entanglement sudden death.  Further
analysis using the evolution of the field $W-$function showed us that for a
given field coherent intensity $\overline{n},$ there is an optimum value of
the decoherence parameters for which the quasiprobability distribution has
no negative values. Given the general interest in superconducting qubits
as a device for quantum information processing, including the realization
of complex single-qubit manipulation schemes and the generation of entangled
states\cite{yam03}, we feel these results may find great utility in future applications.

\begin{acknowledgments}
We gratefully acknowledge Jason Ralph and M A Bouchene for truly
enlightening discussions and useful criticism of this work. MA
gratefully acknowledges the financial support from Universit Paul
Sabatier.
\end{acknowledgments}

\end{document}